\documentclass[12pt,preprint]{aastex}

\shorttitle{The high energy spectrum of PKS 2155$-$304}
\shortauthors{F. Aharonian, A. Abdo, et al. (H.E.S.S. and {\it Fermi}-LAT collaborations)}

\begin{document}

\title{Simultaneous Observations of PKS 2155$-$304 with 
       H.E.S.S., {\it Fermi}, {\it RXTE} and ATOM: 
Spectral Energy Distributions and Variability in a Low State}

\author{ \small The H.E.S.S. Collaboration\\F. Aharonian\altaffilmark{1a,13a},
A.G.~Akhperjanian \altaffilmark{2a},
G.~Anton \altaffilmark{16a},
U.~Barres de Almeida \altaffilmark{8a,30a},
A.R.~Bazer-Bachi \altaffilmark{3a},
Y.~Becherini \altaffilmark{12a},
B.~Behera \altaffilmark{14a},
K.~Bernl\"ohr \altaffilmark{1a,5a},
C.~Boisson \altaffilmark{6a},
A.~Bochow \altaffilmark{1a}
% ...
}


\altaffiltext{1a}{
Max-Planck-Institut f\"ur Kernphysik, P.O. Box 103980, D 69029
Heidelberg, Germany}
\altaffiltext{2a}{
 Yerevan Physics Institute, 2 Alikhanian Brothers St., 375036 Yerevan,
Armenia}
\altaffiltext{3a}{
Centre d'\'Etude Spatiale des Rayonnements, CNRS/UPS, BP 44346, F-31028 Toulouse Cedex 4, France}
% ...
\altaffiltext{5a}{
Institut f\"ur Physik, Humboldt-Universit\"at zu Berlin, Newtonstr. 15,
D 12489 Berlin, Germany}
\altaffiltext{6a}{
LUTH, Observatoire de Paris, CNRS, Universit\'e Paris Diderot, 5 Place Jules Janssen, 92190 Meudon, 
France}
% ...
\altaffiltext{8a}{
University of Durham, Department of Physics, South Road, Durham DH1 3LE,
U.K.}
% ...
\altaffiltext{12a}{
Astroparticule et Cosmologie (APC), CNRS, Universite Paris 7 Denis Diderot,
10, rue Alice Domon et Leonie Duquet, F-75205 Paris Cedex 13, France
\& UMR 7164 (CNRS, Universit\'e Paris VII, CEA, Observatoire de Paris)}
\altaffiltext{13a}{
Dublin Institute for Advanced Studies, 5 Merrion Square, Dublin 2,
Ireland}
\altaffiltext{14a}{
Landessternwarte, Universit\"at Heidelberg, K\"onigstuhl, D 69117 Heidelberg, Germany}
\altaffiltext{16a}{
Universit\"at Erlangen-N\"urnberg, Physikalisches Institut, Erwin-Rommel-Str. 1,
D 91058 Erlangen, Germany}
% ...
\altaffiltext{30a}{supported by CAPES Foundation, Ministry of Education of Brazil}


\altaffiltext{1b}{National Research Council Research Associate}
\altaffiltext{2b}{Space Science Division, Naval Research Laboratory, Washington, DC 20375}
\altaffiltext{3b}{W. W. Hansen Experimental Physics Laboratory, Kavli Institute for Particle Astrophysics and Cosmology, Department of Physics and Stanford Linear Accelerator Center, Stanford University, Stanford, CA 94305}
% ...
\altaffiltext{5b}{The Oskar Klein Centre for Cosmo Particle Physics, AlbaNova, SE-106 91 Stockholm, Sweden}
\altaffiltext{6b}{Stockholm Observatory, Albanova, SE-106 91 Stockholm, Sweden}
% ...
\altaffiltext{8b}{Laboratoire AIM, CEA-IRFU/CNRS/Universit\'e Paris Diderot, Service d'Astrophysique, CEA Saclay, 91191 Gif sur Yvette, France}
% ...
\altaffiltext{12b}{Dipartimento di Fisica ``G. Galilei", Universit\`a di Padova, I-35131 Padova, Italy}
\altaffiltext{13b}{Department of Physics, Royal Institute of Technology (KTH), AlbaNova, SE-106 91 Stockholm, Sweden}
\altaffiltext{14b}{Department of Physics, Center for Cosmology and Astro-Particle Physics, The Ohio State University, Columbus, OH 43210}
% ...
\altaffiltext{16b}{Dipartimento di Fisica, Universit\`a degli Studi di Perugia, I-06123 Perugia, Italy}
% ...
\altaffiltext{30b}{Universit\'e de Bordeaux, Centre d'\'Etudes Nucl\'eaires Bordeaux Gradignan, UMR 5797, Gradignan, 33175, France}
% ...

\begin{abstract}
We report on the first simultaneous observations that cover the
optical, X-ray, and high energy gamma-ray bands of the BL Lac object
\objectname{PKS 2155$-$304}. The gamma-ray bands were observed for 11
days, between 25 August and 6 September 2008 (MJD$\,54704$--$54715$),
jointly with the {\it Fermi Gamma-ray Space Telescope} and the
H.E.S.S.  atmospheric Cherenkov array, providing the first
simultaneous MeV--TeV spectral energy distribution (SED) with the new generation of $\gamma$-ray telescopes. The ATOM
telescope and the {\it RXTE} and {\it Swift} observatories provided
optical and X-ray coverage of the low-energy component over the same
time period. The object was close to the lowest archival X-ray and
Very High Energy (VHE, $>100\,$GeV) state, whereas the optical flux
was much higher. The light curves show relatively little ($\sim 30\%$)
variability overall when compared to past flaring episodes, but we
find a clear optical/VHE correlation and evidence for a correlation of
the X-rays with the high energy spectral index.  Contrary to previous
observations in the flaring state, we do not find any correlation
between the X-ray and VHE components. Although synchrotron
self-Compton models are often invoked to explain the SEDs of BL Lac
objects, the most common versions of these models are at odds with the
correlated variability we find in the various bands for
\objectname{PKS 2155$-$304}.
\end{abstract}


\section{Introduction}

The underlying particle distributions of blazars are usually studied
by matching broadband observations with predictions from radiative
models. Since these sources are highly variable, simultaneous
observations are essential. The most energetic BL Lac spectra extend
up to TeV energies, and positive detections have usually
indicated flaring states. However, with their improved sensitivity,
the new generation of Atmospheric Cherenkov Telescopes (ACTs), which
has more than quadrupled\footnote{See, e.g., the online TeVCat catalog
  http://tevcat.uchicago.edu, which has 22 sources at the time of the
  writing of this Letter.} the number of known extragalactic VHE
sources, finds a few of these sources in marginally variable states
with consistent detections after short exposures. One of these
objects, the blazar \objectname{PKS 2155$-$304} at $z=0.116$, is an
ideal target for such studies. Crucial information is expected from
the {\it Fermi} Gamma-ray Space Telescope, since its improved
sensitivity over EGRET would constrain dramatically the existing
models that predict a wide variety of fluxes in the $100\,{\rm
  MeV}$--$10\,{\rm TeV}$ energy range. Since the H.E.S.S. experiment
detects this source in a low state within $\sim 1\,{\rm h}$,
significant daily detections were guaranteed and the source was
targeted for a 11-day multiwavelength campaign.

\section{Observations and Analysis Results}


The H.E.S.S. observations of \objectname{PKS 2155$-$304} took
place during MJD$\,54701$--$54715$, for a total of 42.2 hours. After
applying the standard H.E.S.S. data-quality selection criteria,
an exposure of 32.9 hours live time remains (MJD$\,54704$--$54715$),
at a mean zenith angle of $18.3^{\circ}$. The data set has been
calibrated using the standard H.E.S.S. calibration method
\citep{aha04}. The analysis tools and the event-selection criteria
used for the VHE analysis are presented in \cite{pks2155_2005-2006}.
The events have been selected using ``loose cuts'', preferred for
their lower energy threshold of $200\,\mathrm{GeV}$ and higher
$\gamma$-ray acceptance. A $0.2^{\circ}$ radius
circular region centered on \objectname{PKS 2155$-$304} was defined to
collect the on-source events. The background was estimated using the
``Reflected Region'' method \citep{aha06}. Those observations yield
an excess of $8800$ events, a signal with a significance of $55.7
\sigma$ calculated following \cite{LiMa83}. Using standard cuts an
excess of $3612$ events with a significance of $68.7 \sigma$ is
found. An independent analysis and calibration
\citep{ben05} yields similar results.

The data from the Large Area Telescope (LAT; \citealt{atw08}) have
been analyzed by using {\tt ScienceTools v9.7}, which will be
publicly available from the {\it HEASARC} in the future. Events having
the highest probability of being photons (class 3, called ``diffuse'')
and coming from zenith angles $< 105^\circ$ (to avoid Earth's albedo)
were selected. The diffuse emission along the plane of the Milky Way,
mainly due to cosmic-ray interactions with the Galactic interstellar
matter, has been modeled using the {\tt 54\_59Xvarh7S} model prepared
with the {\it GALPROP} code \citep{str04a,str04b} which has been
refined with {\it Fermi}-LAT data taken during the first 3 months of
operation. The extragalactic diffuse emission and the residual
instrumental background have been modeled as an isotropic power-law
component and included in the fit. Photons were extracted from a
region with $10^\circ$ radius centered on the coordinates of
\objectname{PKS 2155$-$304} and analyzed with an unbinned maximum
likelihood technique \citep{cas79,mat96} using the {\tt Likelihood}
analysis software provided by the LAT team. Because of calibration
uncertainties at low energies, data in the 0.2--300\,GeV energy band
were selected.

A total of $75\,{\rm ks}$ of exposure was taken with {\it RXTE},
spread over 10 days coinciding with the H.E.S.S. observations, and a
$6.4\,{\rm ks}$ exposure with {\it Swift} was made towards the end of
the campaign. The data taken with the PCA (\citealt{jah95}) and
the XRT (\citealt{bur05}) instruments were
analyzed using the {\it HEASOFT 6.5.1} package using the Guest
Observer Facility recommended criteria. The XRT data were
extracted from a $56\arcsec$ slice, both for the source and the
background. Since the rate was less than 10 Hz, no pile-up is expected
in the Windowed Timing (WT) mode.

During the multiwavelength campaign a total of 106 observations were
taken with the $0.8\,{\rm m}$ ATOM optical telescope \citep{hau04} located on
the H.E.S.S. site. Integration times between $60\,{\rm s}$ and $200\,{\rm s}$ in
the Bessel {\it BVR} filter bands were used. Photometric accuracy is
typically between $0.01$ and $0.02\,{\rm mag}$ for {\it BVR}.

\subsection{Spectral Analyses}
\label{spect}

The H.E.S.S. time-averaged photon spectrum is derived using a
forward-folding maximum likelihood method \citep{spectremethod}. The
very high energy data are well described by a power-law of the form
$dN/dE =I_0(E/E_0)^{-\Gamma}$, with a differential flux at $E_0 =
350\,$GeV (the fit decorrelation energy) of $I_0=10.4\pm0.24_{\mathrm{stat}}\pm \, 2.08_{\mathrm{sys}}
\times 10^{-11}\, \mathrm{cm}^{-2}\, \mathrm{s}^{-1}\,
\mathrm{TeV}^{-1}$ and a spectral index $\Gamma=3.34 \pm
0.05_{\mathrm{stat}} \pm 0.1_{\mathrm{sys}}$. As before, during
non-flaring states of PKS\,2155$-$304, the spectrum, measured with
limited event statistics, shows no indication of curvature. The
spectral index is similar to that previously measured by H.E.S.S. when the source was at a comparable flux level, in 2003
\citep{PKS2155_2003,PKS2155_2003_mwl} and between 2003 and 2005
\citep{pks2155_2005-2006}. The VHE spectrum is affected
by interactions with the Extragalactic Background Light (EBL) which
modifies the intrinsic shape and intensity. Using the $P0.45$ model
\citep{aha05}, the intrinsic spectral index is derived to be
$\Gamma_{\rm int} \approx 2.5$.

The average {\it Fermi} spectra over the duration of the campaign are fitted by a simple power
law for which $I_0=(2.42\pm0.33_{\rm stat}\pm 0.16_{\rm sys})\times 10^{-11} {\rm
  cm^{-2}}\,{\rm s^{-1}}\,{\rm MeV^{-1}}$, $\Gamma=1.81\pm0.11_{\rm stat}\pm 0.09_{\rm sys}$,
and $E_0=943\,{\rm MeV}$ is the energy at which the correlation between the fitted values of
$\Gamma$ and $I_0$ is minimized. The total exposure is $7.7 \times 10^8\,{\rm cm^{2}}\,{\rm
  s}$. There is no statistical preference for a broken power law in this data set. The light
curve derived for {\it Fermi} data between ${\rm MJD}\,54682$--$54743$ shows a similar state on
average as during this campaign, so in order to increase the photon
statistics for the spectral fits, those data were included, resulting in an increase of the exposure by a factor
of $3.6$. The longer data set is then fit by a broken power law spectrum, which is preferred over the
single power law with a significance of $97\%$ using the likelihood ratio test. We obtain a
low-energy photon index of $\Gamma_L=1.61\pm0.16_{\rm stat}\pm0.17_{\rm sys}$, a break energy
of $E_{br}=1.0\pm0.3\,$GeV, a high-energy index of $\Gamma_H=1.96\pm0.08_{\rm stat}\pm0.08_{\rm
  sys}$, and a $0.2$--$300\,{\rm GeV}$ flux of $ (1.13 \pm 0.05_{\rm stat}\pm0.11_{\rm sys})
\times 10^{-7}\,{\rm cm^{-2}}\,{\rm s^{-1}}$. The {\it Fermi} spectrum is consistent with the hard photon index of
  $1.71\pm0.24$ during a flaring episode detected by EGRET \citep{ves95}, but it differs from
  the Third EGRET Catalog spectrum \citep{har99} where the index is $2.35\pm0.26$.

The $4$--$10\,{\rm keV}$ PCA and $0.5$-$9\,{\rm keV}$ XRT
data were analyzed simultaneously with {\tt XSPEC v12.4.0}
\citep{arn96}, using a broken power law model and taking into account the
uncertainty in the cross-calibrations, as well as the variability
across the non-simultaneous observations, by using a
multiplicative factor for each instrument (fixed to 1 for the PCA
data) as in \citet{fal06}. Using a fixed Galactic hydrogen column of
$N_{\rm H}=1.48\times10^{-20}\,{\rm cm^{-2}}$, we obtain a low-energy
photon index of $\Gamma_1=2.36\pm0.01$, a break energy of
$E_{br}=4.44\pm0.48\,{\rm keV}$, and a high-energy index of
$\Gamma_2=2.67\pm0.01$, for an unabsorbed $2$--$10\,{\rm keV}$ flux of
$4.99\times 10^{-11}\,{\rm ergs}\,{\rm cm^{-2}}\,{\rm s^{-1}}$, which
is approximately 2 times higher than during the 2003 campaign
\citep{aha05}. This is similar to the VHE flux increase reported
above, while still being well below the high state fluxes reported by \cite{ves95}.

\subsection{Light Curves}

The light curves from H.E.S.S., {\it Fermi}, {\it RXTE} and ATOM are shown in Fig.~\ref{lightcurve}, where the H.E.S.S.
runs ($\sim 28\,{\rm min}$) were combined to derive nightly flux
values. The average integrated flux above 200 GeV, $(5.56 \pm
0.13_{\mathrm{stat}} \pm 1.11_{\mathrm{sys}}) \times
10^{-11}\,\mathrm{ph\,cm}^{-2}\,\mathrm{s}^{-1}$, corresponds to $\sim
20 \% \, \mathrm{F}_{\mathrm{Crab}>200\mathrm{GeV}}$, or $\sim50\%$
higher than the quiescent state of 2003 \citep{aha05} and 70 times
lower than its peak flaring flux \citep{aha07}. The positive excess
variance $\sigma^2_{\rm XS}$, indicating variability, allows a
fractional root mean square (rms) of $F_{\rm var,VHE}=23 \pm 3\%$ (see
\citealt{vau03} for definitions of $\sigma^2_{\rm XS}$ and
$F_{\rm var}$) to be derived, which is 3 times less than the high
state flaring variability reported by \cite{aha07}. A spectrum was
obtained for each night when possible, otherwise two or three nights
were combined. No indication of spectral variability was found during
those observations, with a limit on the nightly index variations of
$\Delta\Gamma<0.2$.

The {\it Fermi} light curve shows the photon fluxes for the high energy (HE)
range, 0.2--$300\,$GeV, and the photon spectral indices for each
interval. Each bin is the result of a power law fit, using the
background values found on the overall time-averaged fit, and centered
on the H.E.S.S. observations. The light curve fit to a constant has a $\chi^2$ probability of
$p(\chi^2)=0.95$, clearly consistent with a constant flux. The normalized excess variance of $-0.16\pm0.09$
sets a 90\% confidence level limit of $F_{\rm var, HE} \leq 20 \%$ on the fractional variance \citep{fel98}.

The X-ray light curve, derived from spectral fits of the nightly \textit{RXTE}
(and \textit{Swift}) data sets, shows flux doubling episodes on time scales of
days, similar to the optical and VHE measurements. The lowest fluxes
of $\sim 3$--$6\times 10^{-11}$erg\,cm$^{-2}$s$^{-1}$ are at the same
level as those seen in the low state \citep{PKS2155_2003_mwl} but with larger
fluctuations, $F_{\rm var,X}=35 \pm 0.05 \%$. The time history of the fitted
spectral indices in Fig.~\ref{lightcurve} show clearly that the X-ray spectrum
hardens significantly, $\Delta\Gamma_x \approx 0.5$, as the 2--10\,keV
flux increases.

The ATOM fluxes are $\sim 5$ times higher than the low state
found in \cite{PKS2155_2003_mwl}, but the $V$-band magnitudes reported here are
in the range $12.7$--$13$ which is well on the lower side of the
measurements of \objectname{PKS 2155$-$304} reported by \cite{fos08}
when the source was quoted to be in a low state with $V$-band magnitudes in the range $12$--$12.7$. The host galaxy flux
is estimated to be $\approx 10^{-11}\,{\rm ergs}\,\,{\rm cm^{-2}}\,{\rm
  s^{-1}}$ \citep{KS98}, hence most of the optical flux can be
attributed to the central AGN. The average fractional rms over all
bands is $F_{\rm var, opt}\sim 8 \pm 0.5 \%$. The $B-R$ lightcurve
is compatible with a constant, $p(\chi^2)=0.66$, indicating little or no optical spectral
variability.

\section{Discussion}
\label{discussion}

The two-component broad band spectra of high energy-peaked BL Lac objects (HBLs) are typically
modeled with synchrotron self-Compton (SSC) scenarios (e.g., \citealt{ban85}). Despite the
simplicity of these models, they have been successful in reproducing many blazar SEDs and make
definite predictions for the flux and spectral variability that should be seen in the two
components. In particular, for typical parameters, the electrons responsible for the X-ray
emission also produce the VHE emission; and if the underlying particle distributions were to
vary, the resulting flux and spectral changes in the VHE band should be related to variations
in the X-rays. In fact, for the July 2006 flare, a non-linear relationship was seen between the
X-ray and VHE bands, though the observed variability patterns do not quite fit the simple SSC
model in detail \citep{cos08}.

\begin{figure}
\includegraphics[width=\linewidth]{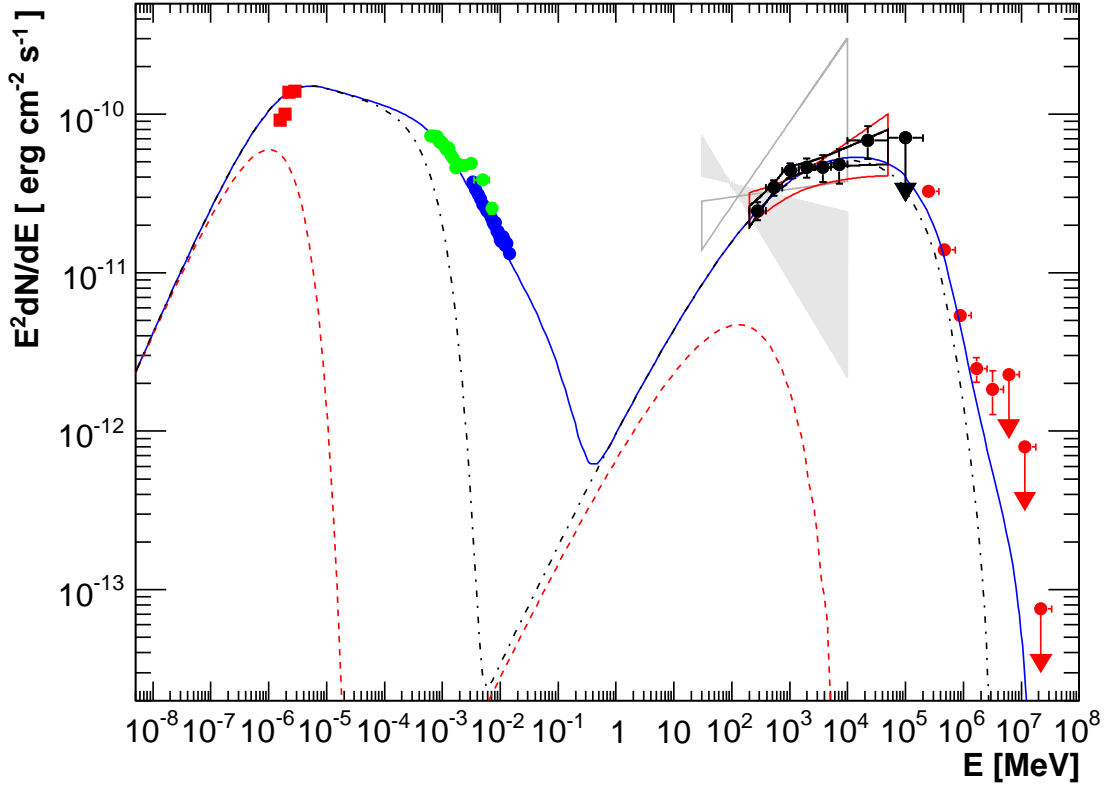}
\caption{The SED of PKS~2155$-$304. The red butterfly is the {\it Fermi} spectrum restricted to
  the ${\rm MJD}\, 54704$--$54715$ period, while the black butterfly covers ${\rm
    MJD}\,54682$--$54743$. As a cross check of the fit robustness, the differential flux was
  estimated in 8 limited energy bins by a power law fit (black circles) and are found to be
  consistent within $1\sigma$ of the global fit, including a clear spectral break at
  $\sim1\,{\rm GeV}$. The gray points are archival NED data, and the two gray butterflies are
  EGRET measurements. The solid
  line is a 1-zone SSC model. The dashed and the dot-dashed lines are the same model without
  electrons above $\gamma_1$ and $\gamma_2$, respectively. The VHE part is absorbed with the
  $P0.45$ extragalactic background model described in \cite{aha05}.}
\label{SED}
\end{figure}

In Fig.~\ref{SED}, we overlay a model SSC calculation that roughly
fits the time-averaged SED. The electron distribution model
parameters, a three-component power-law with indices $p_0 = 1.3$, $p_1
= 3.2$, $p_2 = 4.3$ ($dn/d\gamma \propto \gamma^{-p_i}$), minimal and
maximal Lorentz factors $\gamma_{min}=1$ and $\gamma_{max}=10^{6.5}$,
break electron Lorentz factors $\gamma_1 = 1.4\times 10^4$, $\gamma_2
= 2.3 \times 10^5$, and total electron number $N_{\rm tot} = 6.8\times
10^{51}$, have been set to reproduce the shape of the lower energy
component of the SED. The overall SED is then adjusted with the
remaining parameters: radius of the emitting region in the comoving
frame, $R = 1.5 \times 10^{17}\,$cm; bulk Doppler factor, $\delta =
32$; magnetic field, $B = 0.018\,$G. Even though we regard this fit as
a ``straw-man'' model, it is perhaps reassuring that the joint {\it
  Fermi}-H.E.S.S. time-averaged spectra can be reasonably
well-described as SSC emission. \cite{kat08} found similar values for
$R$, $B$ and $\delta$ in their SSC description of a steady large jet
component in the SED of PKS~2155$-$304.

Some features of this model calculation are particularly noteworthy.
The electrons that produce the synchrotron X-ray emission have Lorentz factors
$>\gamma_2$. When the power-law component for those electrons is
omitted from the calculation, the dot-dashed curve in Fig.~\ref{SED}
results. For this particular set of parameters, the electrons that
produce the X-rays have higher energies than the electrons that
produce the VHE emission. Furthermore, the lack of a significant
impact on the shape of the SSC component when those electrons are
removed indicates that Klein-Nishina effects suppress any significant
contribution by those electrons to the emission at $\sim\,$TeV
energies.  

\begin{figure}
\includegraphics[width=\linewidth]{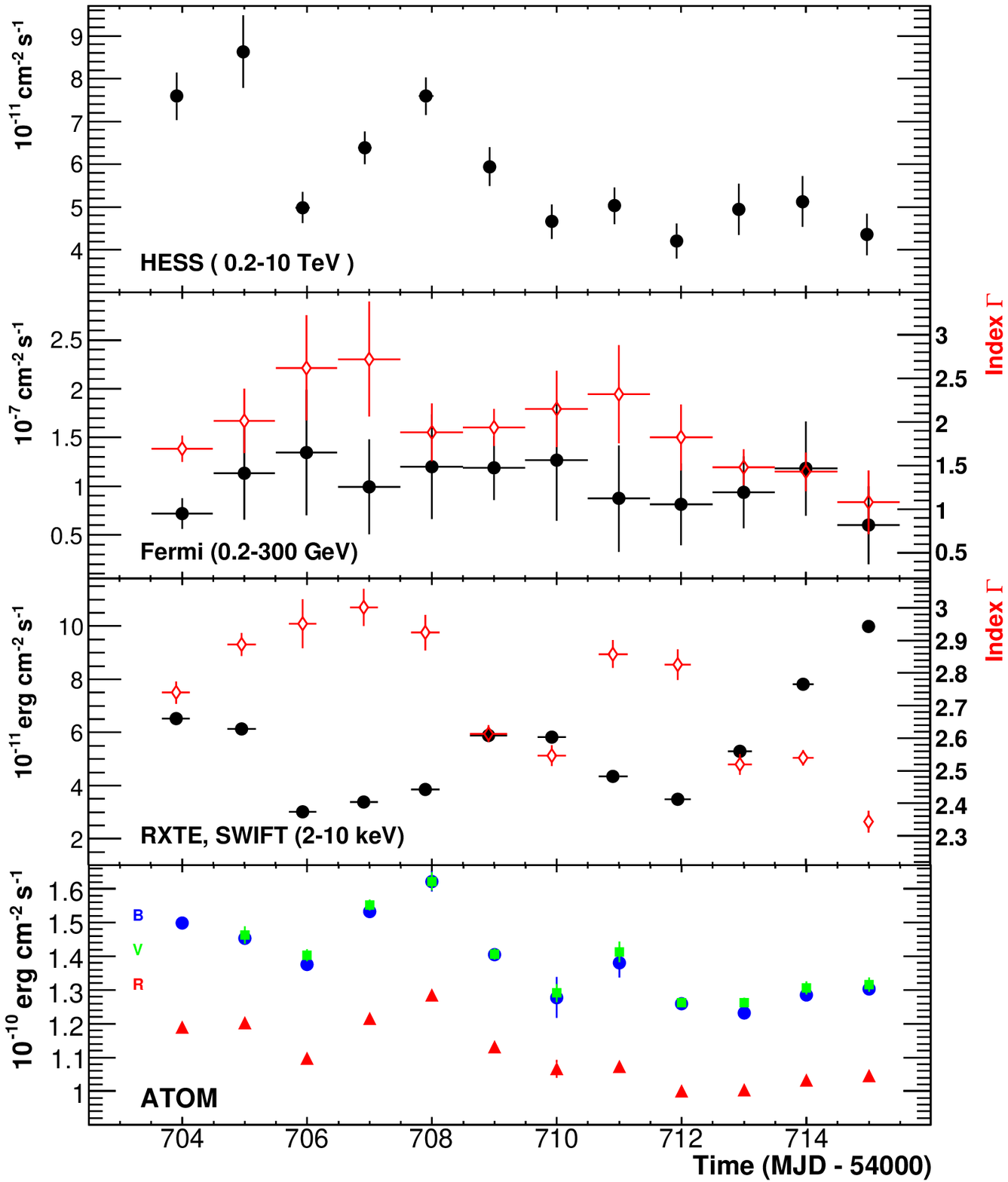}
\caption{Light curves from (top to bottom): H.E.S.S., {\it
    Fermi}, {\it RXTE/Swift}, and ATOM. The {\it Fermi} and
  {\it RXTE/Swift} panels also show the spectral index measurements (red) for
  each night. Vertical bars show statistical errors only. Horizontal
  bars represent the integration time and are apparent only for the {\it RXTE}
 and {\it Fermi} data. The ATOM bands are $B$ (blue circles), $V$
  (green squares) and $R$ (red squares). }
\label{lightcurve}
\end{figure}

These features of this calculation allow that there need not be a correlation between
the X-ray and VHE fluxes; and in fact, this is what is observed. In contrast with the July
2006 flare, we do not find any evidence of flux correlation between the X-ray and
H.E.S.S. bands with a Pearson's $r$ of $0.12\pm0.1$ between these bands. Furthermore, the
2--10 keV X-ray spectra show spectral variability consistent with an underlying electron
distribution for which the cooling time scales are of order the flux variability time scales,
i.e., the spectra are softer when the flux is lower, with changes in photon index of
$\Delta\Gamma_x \approx 0.5$ (Fig.~\ref{lightcurve}); whereas the VHE emission shows no
evidence for significant spectral variability despite flux variations of a factor of $2$.
Since radiative cooling time scales vary inversely with electron energy, this supports the
conclusion that the electrons responsible for the synchrotron emission in the X-ray band have
higher energies than the electrons that produce the inverse-Compton emission in the VHE range,
assuming they are part of the same overall non-thermal distribution.

Even though this all fits in with our straw-man SED calculation, the
variability patterns in the optical, X-ray, HE and VHE bands suggest a
much more complex situation. In the absence of spectral variability,
the mechanisms that would produce the observed flux variability in the
VHE band are rather constrained. Increases in flux could be driven by
injection of particles with a constant spectral shape, and decreases
in flux could be caused by particle escape from the emitting region or
by expansion (``adiabatic'') losses, assuming those latter two
processes can operate independent of particle energy.

However, since the electrons that produce the VHE emission must be in
the weak radiative cooling regime, a more natural mechanism for the
flux variability would be that changes in the seed photon density are
driving the variability. Comparing the daily flux values in the
optical and the VHE bands, we find indications of fairly strong
correlations that suggest that the optical emission provides the
target photons for the IC emission. In the $B$, $V$, and $R$ bands,
the correlations with the H.E.S.S. fluxes have Pearson's $r$ values in
the range 0.77--0.86 with uncertainties $\le 0.09$. This correlated
behavior is readily apparent in the light curves shown in
Fig.~\ref{lightcurve}, and these results provide the first
quantitative evidence of correlated variability between the optical
and VHE bands on these time scales for an HBL.\footnote{\cite{don08}
  mention possible correlated variability in the recent June 2008
  flare of Mrk~421 in a high state.}  Confirmation of this behavior,
not only from this source but also from other VHE emitting blazars in a low
state, would provide important constraints on emission models for
these objects.

In the context of a single-zone SSC model, we would expect that any
flux variability in the optical bands should also appear as
variability in the {\it Fermi}-LAT energy range. To illustrate this,
we plot, as the dashed curve in Fig.~\ref{SED} the SED that
results if we omit contributions from electrons with energies $>
\gamma_1$. For the original model parameters, the electrons that
produce the optical-soft X-ray emission also produce the bulk of the
IC component, including the HE and VHE emission. Since we do not find
any indication of a correlation between the optical and HE fluxes,
this suggests that the optical emission may arise from a separate
population of electrons than those responsible for the HE and VHE
emission. If so, then these electrons probably also occupy a
distinct physical region with different physical parameters (magnetic
field, size scale, bulk Lorentz factor). Multizone SSC models of this
kind have already been proposed to account for the ``orphan''
$\gamma$-ray flare in 1ES~1959+650 during May 2002 \citep{kra04}.

Although the $0.2$--$300\,$GeV photon fluxes measured by {\it Fermi}
are consistent with being constant, we find more significant variations
of the photon spectral index in the daily analyses ($p(\chi^2)=0.19$). The fitted values
range from fairly soft, $\Gamma = 2.7 \pm 0.7$, to extremely hard,
$\Gamma = 1.1\pm 0.4$.  

These values, along with the constant, intrinsic VHE index of
$\Gamma_{\rm VHE} \approx 2.5$ derived from the H.E.S.S. data,
imply spectral breaks between the HE and VHE bands of $\Delta\Gamma$ 
as large as 1.4. Very sharp spectral breaks ($\Delta\Gamma\ga 1$) would
require rather narrow electron distributions and would therefore pose
difficulties in fitting a broad lower energy component in the context
of a single-zone model. Interestingly, we find a significant
anticorrelation between the nightly X-ray fluxes and the {\it Fermi}-LAT
spectral indices of $r_{X\Gamma} = -0.80 \pm 0.15$. A fit to a linear model is preferred
over a constant at the $2.6\sigma$ level, with a slope of $-0.14 \pm 0.05$. If the electrons
that produce the X-rays are at higher energies than those that produce
the TeV emission, the cause for such a correlation would be difficult
to understand. An important caveat in considering these results is
that the {\it Fermi} coverage for PKS 2155$-$304 was relatively uniform
over each 24 hour period, whereas the optical, X-ray, and VHE
observations were restricted to 4--6 hour intervals each night.
Hence, the {\it Fermi} observations are not strictly simultaneous with
the other measurements, so it is possible that some of the observed HE
spectral variability occurred outside of the nightly observing
windows.

As the first multiwavelength campaign of an HBL that includes {\it
  Fermi} and an ACT instrument, these observations have yielded
results that strongly challenge the standard models for these sources.
Having caught PKS 2155$-$304 in a low state, we see that its spectral
and variability properties are significantly different than its
flaring, high state behavior. The variability patterns, in
particular, defy easy explanation by the usual SSC models and should
provide valuable constraints for models that attempt to describe the
emission mechanisms in blazar jets.

\acknowledgements


{\small The support of the Namibian authorities and of the University of Namibia
in facilitating the construction and operation of H.E.S.S. is gratefully
acknowledged, as is the support by the German Ministry for Education and
Research (BMBF), the Max Planck Society, the French Ministry for Research,
the CNRS-IN2P3 and the Astroparticle Interdisciplinary Programme of the
CNRS, the U.K. Science and Technology Facilities Council (STFC),
the IPNP of the Charles University, the Polish Ministry of Science and 
Higher Education, the South African Department of
Science and Technology and National Research Foundation, and by the
University of Namibia. We appreciate the excellent work of the technical
support staff in Berlin, Durham, Hamburg, Heidelberg, Palaiseau, Paris,
Saclay, and in Namibia in the construction and operation of the
equipment.

The {\it Fermi}-LAT Collaboration acknowledges generous ongoing
support from a number of agencies and institutes that have supported
both the development and the operation of the LAT as well as
scientific data analysis. These include the National Aeronautics and
Space Administration and the Department of Energy in the United
States, the Commisariat \`a l'Energie Atomique and the Centre National
de la Recherche Scientifique / Institut National de Physique
Nucl\'eaire et de Physique des Particules in France, the Agenzia
Spaziale Italiana and the Istituto Nazionale di Fisica Nucleare in
Italy, the Ministry of Education, Culture, Sports, Science and
Technology (MEXT), High Energy Accelerator Research Organization (KEK)
and Japan Aerospace Exploration Agency (JAXA) in Japan, and the
K.~A. Wallenberg Foundation, the Swedish Research Council and the
Swedish National Space Board in Sweden.

Additional support for science analysis during the operations phase
from the following agencies is also gratefully acknowledged: the
Istituto Nazionale di Astrofisica in Italy and the K.A. Wallenberg
Foundation in Sweden for providing a grant in support of a Royal
Swedish Academy of Sciences Research fellowship for JC.}

\clearpage

\end{document}